\documentclass[
    prd, twocolumn,
    superscriptaddress, nofootinbib, amsmath, amssymb,
    aps, floatfix
]{revtex4-1}
\usepackage{color,amsmath,amssymb,graphicx,latexsym,subfigure}
\usepackage{threeparttable,multirow}

\newcommand{\beq}{\begin{equation}}
\newcommand{\eeq}{\end{equation}}
\newcommand{\bea}{\begin{eqnarray}}
\newcommand{\eea}{\end{eqnarray}}
\def\ton{t_\text{on}}
\DeclareMathOperator\erf{erf}

\usepackage[arrows]{ezedits}

\defineEdit{AAA}{\color[rgb]{0.0,0.0,1.0}}{\color[rgb]{0.0,0.0,1.0}}

\begin{document}

\title{Listening for echo from the stimulated axion decay with the 21 CentiMeter Array}

\author{Ariel Arza}
\affiliation{Department of Physics and Institute of Theoretical Physics, Nanjing Normal University, Nanjing 210023, China}

\author{Quan Guo}
\affiliation{Shanghai Astronomical Observatory, Chinese Academic Science, Shanghai 200030, China}

\author{Lei Wu}
\email{leiwu@njnu.edu.cn}
\affiliation{Department of Physics and Institute of Theoretical Physics, Nanjing Normal University, Nanjing 210023, China}

\author{Qiaoli Yang}
\affiliation{College of Physics and Optoelectronic Engineering,
Department of Physics, Jinan University, Guangzhou 510632, China}

\author{Xiaolong Yang}
\email{yangxl@shao.ac.cn}
\affiliation{Shanghai Astronomical Observatory, Chinese Academic Science, Shanghai 200030, China}
\affiliation{Shanghai Key Laboratory of Space Navigation and Positioning Techniques, Shanghai 200030, China}

\author{Qiang Yuan}
\email{yuanq@pmo.ac.cn}
\affiliation{Key Laboratory of Dark Matter and Space Astronomy, Purple Mountain Observatory,
Chinese Academy of Sciences, Nanjing 210023, China}
\affiliation{School of Astronomy and Space Science, University of Science and Technology of China,
Hefei 230026, China}

\author{Bin Zhu}
\email{zhubin@mail.nankai.edu.cn}
\affiliation{Department of Physics, Yantai University, Yantai 264005, China}

\date{\today}

\maketitle

The axion is nowadays one of the leading candidates to compose the dark matter of the universe. A great experimental effort has been performed by the so-called axion haloscopes experiments among others (For complete reviews on updated axion phenomenology, see Ref.~\cite{Irastorza:2018dyq}).

In this letter, we take the new axion dark matter echo idea proposed in Ref.~\citep{Arza:2019nta}. This technique aims to send out of space a powerful beam of electromagnetic radiation in the radio/microwave frequency range and search for the electromagnetic echo produced by stimulated decay of ambient axion dark matter. In this work, we propose to utilize the 21 CentiMeter Array (21CMA) observatory in China to probe axion echo signals in the frequency range $50-200~\text{MHz}$, corresponding to axion masses in the range $4.13\times10^{-7}~\text{eV}<m_a<1.65\times10^{-6}~\text{eV}$~\footnote{The radio astronomy frequency window (where the sky is transparent to radio/microwaves) ranges, approximately, from  30 MHz to 30 GHz, which corresponds to axion masses within the range $2.5\times10^{-7}-2.5\times10^{-4}~\text{eV}$.}. We aim to employ a 1MW emitter,
assuming a total integration time of 2 years to cover the whole 21CMA frequency range. In contrast with the proposals for astrophysical observations \cite{Buen-Abad:2021qvj,Sun:2021oqp,Ghosh:2020hgd}, and optical experimental setups \cite{Beyer:2021mzq,Gong:2023ilg}, our proposal works as an option for lower axion masses and it is also more sensitive up to about one order of magnitude in the axion-photon coupling.

\begin{figure*}[htbp]
\centering
\includegraphics[width=1\textwidth]{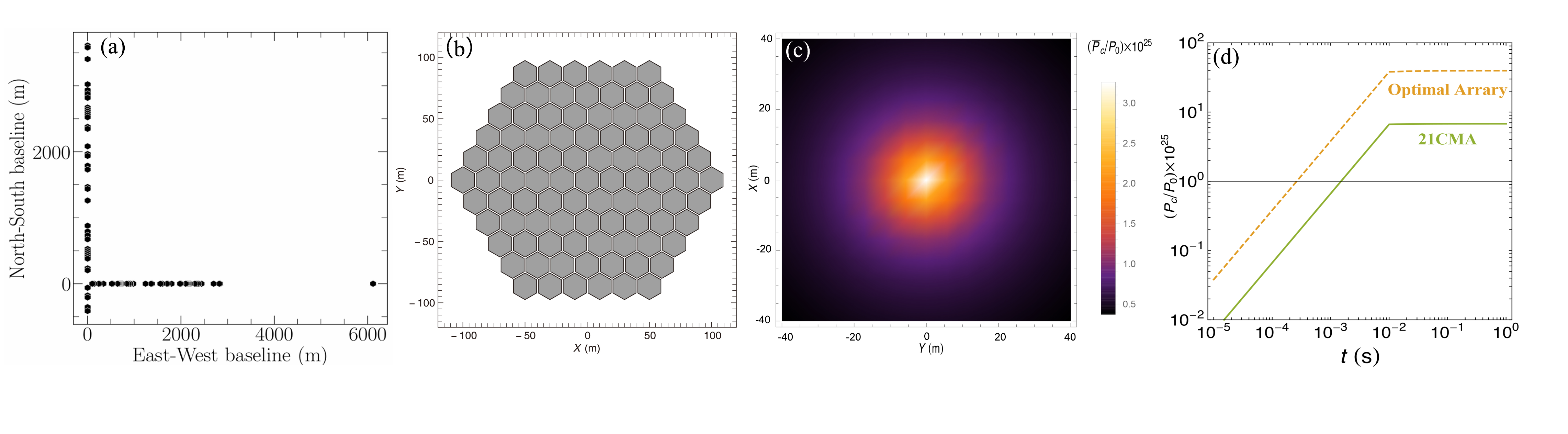}
\caption{(1) the current layout of the 21CMA array. (2) optimized layout 
(with additional 10 pods) for the axion echo experiment. 
(3) spatial distribution of the echo signal integrated over $\ton$.
(4) axion echo time profiles for the current 21CMA layout (dashed) 
and the optimized layout (solid). (Color online)}
\label{fig:signal}
\end{figure*}

The 21CMA is a unique low-frequency radio interferometer, which is located in the Tianshan Mountains in western China\footnote{\url{http://21cma.bao.ac.cn}}. This ground-based meter-wave interferometric array is designed to probe the 21 cm radiation of neutral hydrogen from the cosmic dawn and the epoch of reionization at $z=6\sim27$. It was constructed from August 2005 to July 2006 and upgraded by July 2010. The array consists of 81 pods with 127 log-periodic antennas each, deployed along two perpendicular arms of $6\times4$ km in length (see the first panel of Fig.~\ref{fig:signal}). The array images a field of $10\sim100$ square degrees centered on the North Celestial Pole 24 hours per day in the low-frequency range of 50 MHz to 200 MHz with a resolution of 24 kHz \citep{2016RAA....16...36H, 2016ApJ...832..190Z}.
To optimize the detection of the axion echo, we also consider a re-arrangement of the 21CMA layout to a hexagon shape (with an addition of 10 pods to make a symmetric shape), as shown in the second panel of Fig.~\ref{fig:signal}.

The third panel of Fig.~\ref{fig:signal} shows the spatial distribution of the echo signal averaged over 1 second. The distribution features the expected circular shape and follows a $1/r$ behavior. The fourth panel of Fig.~\ref{fig:signal} presents the time dependence of the power signal $P(t)$, for an emission time of $\ton=1~\text{s}$, and the layouts of the current 21CMA (dashed) and the optimized re-arrangement (solid). To get these profiles, we have assumed a signal frequency of $\nu=100~\text{MHz}$ and an axion-photon coupling $g=10^{-10}~\text{GeV}^{-1}$. The arms of the 21CMA have a length of a few kilometers, so the signal reaches a saturation value after a time of $t_{\rm sat}\sim$ tens of milliseconds. If $\ton<t_{\rm sat}$, the signal would start to decline before reaching the saturated value. The results further show that the optimization of the array layout gives a factor of $\sim5.8$ enhancement of the echo power, due to a higher collection efficiency of the optimized layout for an isotropic signal distribution. The details of the signal computation are given in Sec.~I of Supplementary Material.

In the laboratory frame, the outgoing photon frequency $\nu$ required for stimulated axion decay, and also the echo frequency $\nu_e$, is given by $\nu\approx\nu_e\approx m_a/(4\pi)$. The echo frequency is shifted with respect to the outgoing frequency by the small value $1.5\times10^{-3}\nu$ if we consider the standard halo model (SHM). The signal will suffer contamination from man-made radio frequency interference (RFI) and attenuation from the atmosphere and ionosphere. The detailed analyses and removal methods are discussed in Sec.~II of Supplementary Material.
To suppress the contamination from the beam and the ionosphere reflection, we use a beam whose bandwidth is narrower than the shifted value. In this case, the echo spectrum is also a narrow line with bandwidth equaling the beam bandwidth. As a benchmark, we take\footnote{This bandwidth is narrower than the current 21CMA resolution of 24 kHz. However, the resolution can be improved easily in the data analysis, without any change of the hardware.} $\delta\nu=10\,\text{kHz}\left(\nu\over100\,\text{MHz}\right)$. Note that in this work, the bandwidth is defined as the full width at half maximum (FWHM). The echo power collected by the 21CMA detectors is then
\begin{align}
P_c&=1.4\times10^{-20}~\text{W}\left(g\over10^{-10}~\text{GeV}^{-1}\right)^2 \nonumber
\\
&\ \ \ \ \ \ \ \ \times\left(100\,\text{MHz}\over\nu\right)\left(P_0\over\text{10~kW}\right) \enspace. \label{eq:power2}
\end{align}

We aim to cover the whole 21CMA frequency range, from 50 MHz to 200 MHz, in approximately two years. To do so, the number of outgoing beam shots that we need to send in one year is $n=\nu/\delta\nu=10^4$, each one with a duration of $\tau=3.16\times 10^3$~s. To suppress the background, we would further employ an on-off technique, i.e., to switch on the beam for $\ton$ and then switch off for another $\ton$. Thus the effective beam sending time for each shot is $\tau_{\rm eff}=1.58\times 10^3$~s.

\begin{figure}[htbp]
\centering
\includegraphics[width=0.45\textwidth]{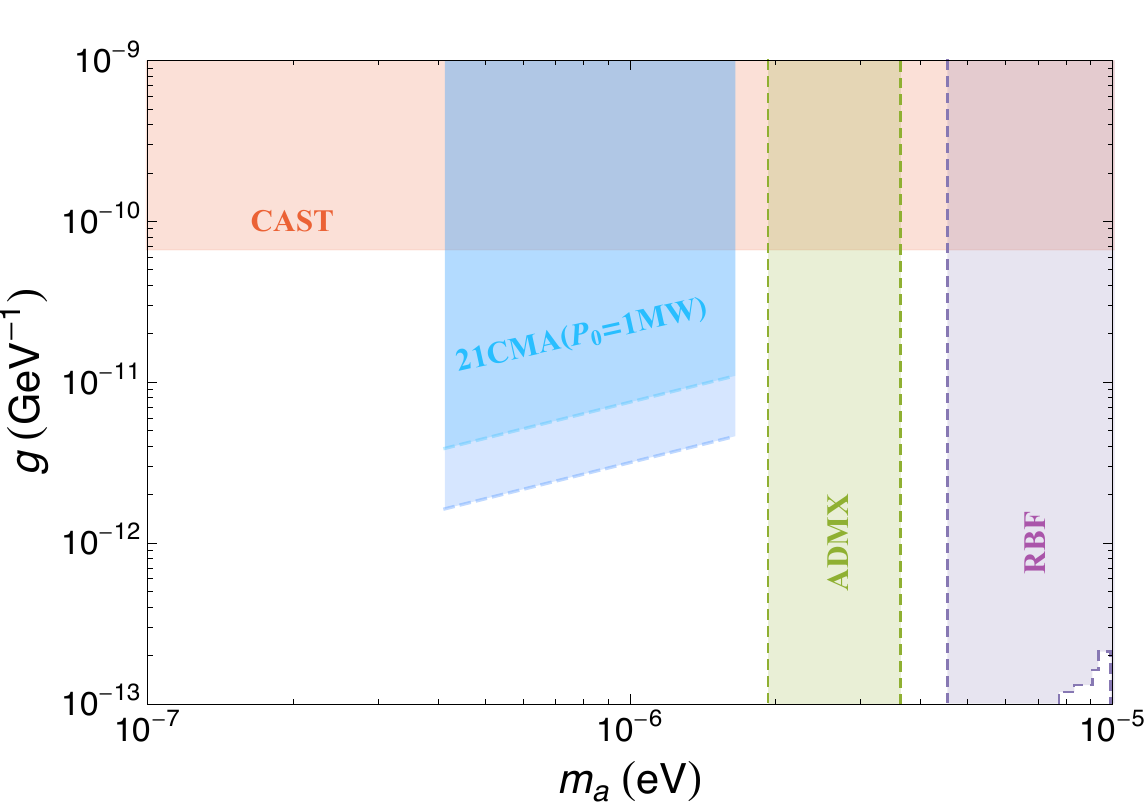}
\caption{Expected 95\% C.L. sensitivity limits on the plane of the axion mass $m_a$ and the axion-photon coupling $g$. We take the power of emitter $P_0=1$ MW. The bounds of CAST, ADMX, and RBF are shown as well \cite{Irastorza:2018dyq}. (Color online)}
\label{fig:space}
\end{figure}

We estimate the achievable sensitivity via a likelihood analysis by setting the outgoing beam 
to a periodic on-off mode. The construction of the likelihood can be found in Sec.~III of Supplementary Material. We generate the simulation data based on the detector's noise for 
the background-only hypothesis. The noise power of the detector array can be estimated as
\begin{equation}
P_{\rm noise}=\frac{2k_BT_{\rm sys}\delta \nu}{\sqrt{2\delta\nu\Delta t}},
\label{noise}
\end{equation}
where $T_{\rm sys}$ is the system temperature, $k_B$ is the Boltzmann constant, and $\Delta t$ 
is the observing time. Here we assume that the receiver resolution equals the echo bandwidth,
which is $10^{-4} \nu$. 
For the 21CMA, $T_{\rm sys}$ is approximately 300 K \citep{2013ApJ...763...90W}, which is the 
total noise temperature with contributions from the receiver, the antenna, the atmosphere, 
and the astrophysical background. Note that, using an on-off observation mode allows the 
quasi-constant part of the noise to be subtracted, leaving a fluctuation noise lower than 
$T_{\rm sys}$, i.e., the thermal noise temperature of the 21CMA receiver which is about 
$T_{\rm th}=50$ K \citep{2016RAA....16...36H, 2016ApJ...832..190Z}. The resulting measurements 
are the power of the blank sky, $P_{i}$, with $rms$ noise $\sigma_{i}$, where $i$ denotes the 
time bin with width $\Delta t_i$. To calculate $\sigma_{i}$, we replace $T_{\rm sys}$ in 
Eq. (\ref{noise}) with $T_{\rm th}=50$ K.  Fig.\ref{fig:space} presents a comparative analysis of various signal power levels and experimental setups, including CAST, 21CMA $(P_0=1\mathrm{MW})$, ADMX, and RBF, showing their respective operational parameters and detection capabilities and highlighting the superior detection capabilities of $21\mathrm{CMA}$ compared to CAST.

The axion is a hypothetical elementary particle that could solve the long-standing strong CP problem in particle physics and the dark matter mystery in the cosmos. By sending radio waves to space, the axion dark matter can be stimulated to generate a reflected signal and thus be detected by terrestrial telescopes. As a pathfinder, we study the expected sensitivity of searching for the axion dark matter between 0.41 and 1.6 $\mu\text{eV}$ with the 21 CentiMeter Array (21CMA). For a 1 MW emitter covering the whole 21CMA frequency range in two years, we find that the resulting sensitivity on the axion-photon coupling surpasses helioscope bounds by about an order of magnitude.

\section*{Author Contributions}
Ariel Arza contributed to signal calculations and the writing of all sections of the paper. Quan Guo contributed to understanding of the 21CMA experimental setup. Lei Wu contributed to the original idea, analytic calculations and writing throughout. Qiaoli Yang contributed to understanding of the signal and the corresponding writing. Xiaolong Yang contributed to numerical calculations, background simulation, and the writing of all sections of the paper. Qiang Yuan contributed to signal calculations, data analysis, and the writing. Bin Zhu contributed to signal calculations, background simulation, and the writing of all sections of the paper. 

\acknowledgments
A.A. thanks Nanjing Normal University for its hospitality while working on part of this paper. This work is supported by the National Natural Science Foundation of China (12275134, 12275232, 12335005, 12147228, 12150010, 12103076), the Project for Young Scientists in Basic Research of Chinese Academy of Sciences (No. YSBR-061), and the Ministry of Science and Technology of China (grant No. 2020SKA0110200).

\pagebreak
\widetext
\begin{center}
\textbf{\large Supplemental Materials: Listening for echo from the stimulated axion decay with the 21 CentiMeter Array}
\end{center}
\setcounter{equation}{0}
\setcounter{figure}{0}
\setcounter{table}{0}
\setcounter{page}{1}

This Supplementary Material includes the calculations of the signal power, the removal methods of possible contamination from the man-made radio frequency interference (RFI) and attenuation from the atmosphere and ionosphere, and the construction of the likelihood in our numerical analysis.

\section{Signal Power}

\begin{figure*}[ht]
\centering \includegraphics[scale=0.1]{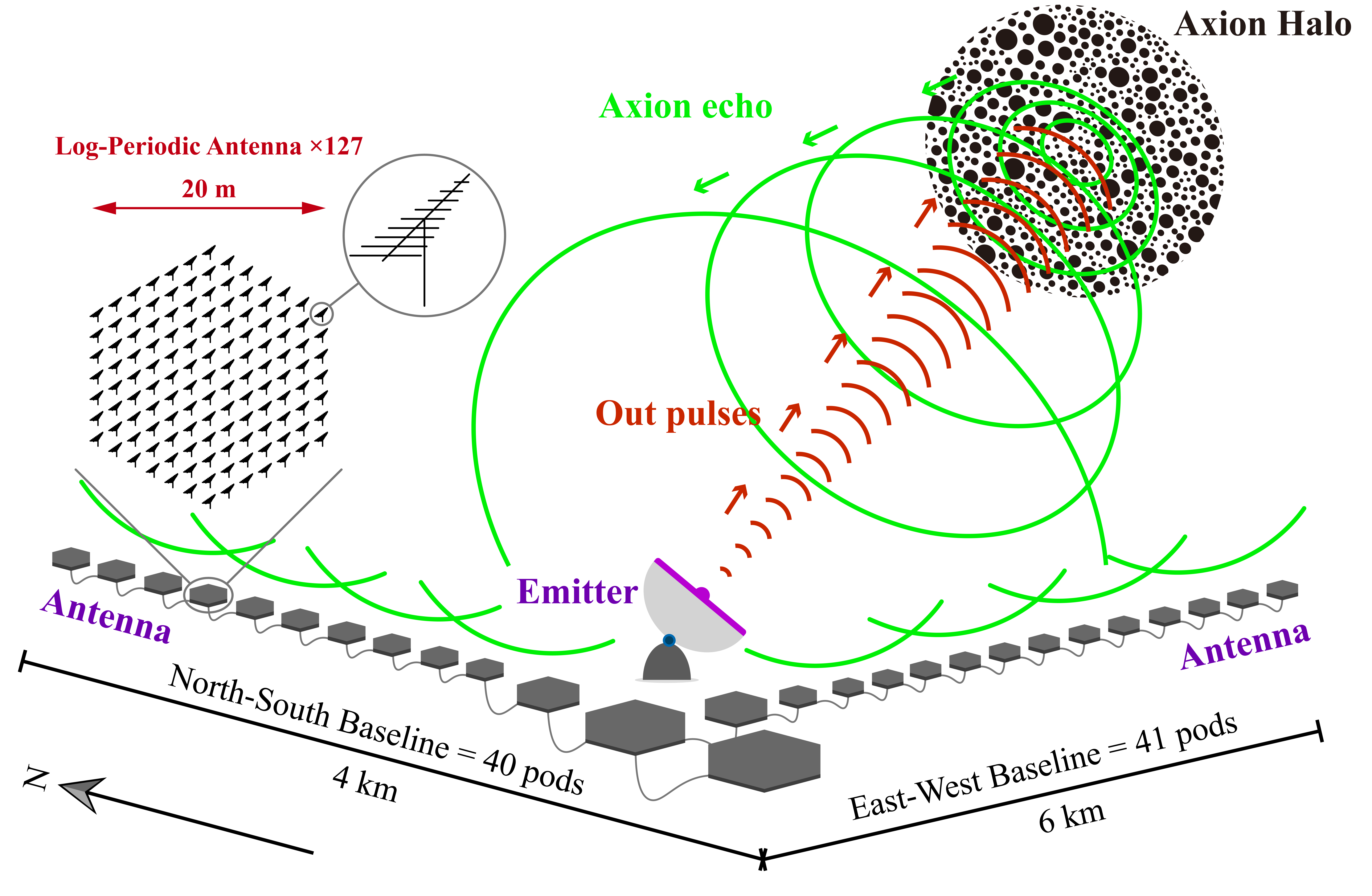}
\caption{Sketch map of using the 21 Centimeter Array in Xinjiang, China, and an emitter to activate axion and capture corresponding echo emissions. The map displays the location of the two primary components of the array: the North-South Baselines and the East-West Baselines. The 21CMA is a radio interferometer designed to study the large-scale structure of the universe and the epoch of reionization. This figure provides a visual representation of the site and its components, which are critical for the success of the 21CMA's scientific goals. In particular, the figure illustrates the potential to detect axion echoes at the 21CMA site by emitting a powerful beam into the sky. (Color online)
\label{fig:sketch_map}}
\end{figure*}

The basic setup of the experiment includes a radio emitter and a receiving antenna array with specific layout (see Fig \ref{fig:sketch_map}). The emitter sends a beam of radio waves with power $P_0$ and physical size (area) $S_0=\pi R_0^2$. The beam is collimated by a dish antenna of radius $R_0$, where we assume a full efficiency of the dish. 

For an emitter turned on during a time $\ton$, the echo intensity, as a function of polar coordinates $\vec r=(r,\theta)$ and time $t$, for a particular axion transverse velocity $\vec v_\perp$ is \cite{Arza:2021nec,Arza:2022dng}
\begin{equation}
\begin{aligned}
&I(r,\theta,v_{\perp},t)={\pi\over32\sqrt{2}}{g^2\rho\over\Delta}{P_0\over S_0}
{R_0\over v_\perp}\exp{\left(\frac{-r^2\sin^2\theta}{R_0^2}\right)} \nonumber
\\
&\times\left\{\begin{array}{ll}
\erf\left(\frac{r\cos\theta}{R_0}\right)+\erf\left(\frac{v_\perp t-r\cos\theta}{R_0}\right) & \text { for } t<\ton \\
\erf\left(\frac{v_\perp t-r\cos\theta}{R_0}\right)-\erf\left(\frac{v_\perp (t-\ton)-r\cos\theta}{R_0}\right) & \text { for } t>\ton
\end{array}\right.,
\end{aligned}
\end{equation}
where $\rho$ represents the local axion energy density and $g$ is the effective axion-photon coupling. Here, the polar coordinates are defined such that $\theta$ is the angle between $\vec r$ and $\vec v_\perp$.

The echo is produced from the interaction of a photon beam with frequency bandwidth $\delta \nu$ and the axion local energy density with momentum dispersion $\delta \vec{p}$. The parameter $\Delta$ accounts for the maximum between these two dispersions, which can be written as $\Delta=\sqrt{(2\pi\delta\nu/2.355)^2+(\delta p_\parallel/2)^2}$, where $\parallel$ refers to the component parallel to the beam wave vector, and the factor 2.355 accounts for the conversion from FWHM to the Gaussian width.

Averaging over the axion transverse velocity distribution $f_a(v_\perp)$, we have 

\begin{align}
\bar{I}(r,t)=&\int_0^\infty v_\perp dv_\perp f_{a}(v_\perp)\int_0^{2\pi} d\theta\,I(r,\theta,v_\perp,t),\label{intensity2}
\end{align}
where we assume the standard halo model (SHM) Maxwell-Boltzmann velocity distribution  
\begin{equation}
f_a(\vec v_\perp)={3\over2\pi\sigma_v^2}e^{-{3v_\perp^2\over2\sigma_v^2}}, 
\end{equation}
with $\sigma_v=270~\text{km/s}$ being the DM velocity dispersion. 

The time profile of the echo power signal collected by a receiving array can be computed as
\begin{align}
P(t)=&\int_0^{\infty}dr\,r\int_0^{2\pi}d\phi\,\bar{I}(r,t)\,w(r,\phi). 
\label{power_t}
\end{align}
Here $w(r,\phi)$ represents the weight of the array layout, which is 1 for spatial locations with an antenna and 0 without the antenna. The spatial distribution of the echo signal averaged within any time interval $T$ is
\begin{align}
F(r)=&\frac{1}{T}\int_0^{T}dt\,\bar{I}(r,t). 
\label{power_r}
\end{align}

\section{Contamination}
Electromagnetic waves suffer contamination from man-made radio frequency interference (RFI) and attenuation from the atmosphere and ionosphere. The long wavelength ($<1$\,GHz) signals are affected by interference from e.g., digital TV, FM, and communication signals of low-Earth-orbiting satellites (e.g., \cite{2015PASA...32....8O}). The site of 21CMA has a good radio environment, which can minimize such contamination for weak signal detection. We can also take some strategies for RFI mitigation, including: (1) placing electromagnetic shielding for each pod to minimize interference from neighbors and the ground; (2) skipping strong and steady RFI affecting channels; (3) removing the instant RFI by using reference antennas; (4) flagging instant RFI in post-processing steps.

\begin{figure*}[htbp]
\centering
\includegraphics[width=1\textwidth]{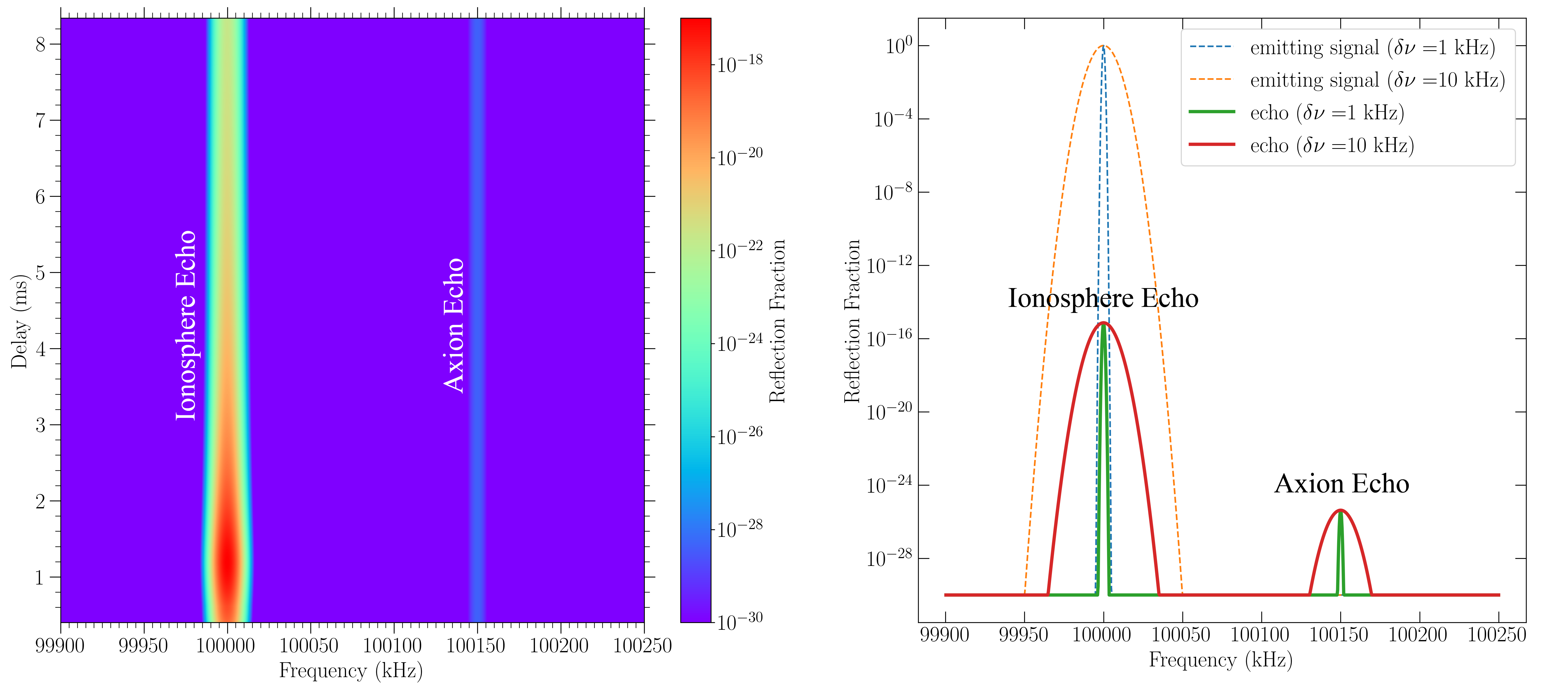}
\caption{Left: the reflection fraction for ionosphere echo and axion echo along with time delay and frequency with a narrow-bandwidth (5\,kHz) signal emitting centered at 100\,MHz. The axion parameters are $g=10^{-12}\,\mathrm{GeV^{-1}}$ and $m_a=1.25~\mu$eV. 
Right: the emitting signals and the integrated reflection fraction for both ionosphere and axion
echos, with the bandwidth of the emitting signal of 1 kHz (green) and 10 kHz (red), respectively. 
Note that the bandwidth is defined as the FWHM of the signal. (Color online)}
\label{fig:narrowband}
\end{figure*}

The ionosphere reflection could be one potential source of contamination. 
By means of the Doppler shift of the axion echo signal (about $10^{-3}\nu$), we can effectively
eliminate the ionosphere reflection contamination. The reflection from the ionosphere is shifted 
with respect to the beam by an amount of $\nu v_\text{ion}/c$, where $v_\text{ion}\sim 100$ m/s 
is the typical group velocity of ionospheric particles. Therefore the frequency shift of the
ionosphere reflection is much smaller than that of the axion echo and can be effectively
distinguished given the beam bandwidth is narrow enough. Fig.~\ref{fig:narrowband} illustrates
the frequency differences among the beam, the ionosphere reflection, and the axion echo. This strategy seems to be very efficient for our purpose, however, there might be some ionospheric particles from the tail of the velocity distribution that are fast enough to reach the echo-shifted frequency. Although we expect this population to be exponentially suppressed, this point deserves to be clarified in the future with a dedicated study.

It is expected that the leakage of the emitter could be a strong source of contamination. The frequency shift can also be used to avoid emitter leakage contamination. For our benchmark setting, a 10 kHz bandwidth could be enough to separate the axion echo signal from the beam signal (see Fig.~\ref{fig:narrowband}). As a reference, the bandwidth coverage for the Sanya incoherent scatter radar site in Sanya, China, ranges from 0.5\,kHz to 1\,MHz \citep{2022JGRA..12730451Y}. It means such a narrow beam is feasible.

\section{$\chi^2$-squared Analysis} \label{signal}

We construct a $\chi^2$ function as
\begin{equation}
\chi^2=\sum_{i}\frac{\left[P_i-P_{{\rm bkg},i}-\bar{P}_{{\rm sig},i}(m_a,g)
\right]^2}{\sigma_i^2},
\end{equation}
where $P_{{\rm bkg},i}$ is the background to be fitted in the analysis, and $\bar{P}_{{\rm sig},i}(m_a,g)$ is the average echo signal in the time bin $\Delta t_i$
\begin{equation}
\bar{P}_{{\rm sig},i}(m_a,g)=\frac{\int_{\Delta t_i}P(t){\rm d}t}{\Delta t_i}.
\end{equation}

\begin{figure}[htbp]
\centering
\includegraphics[width=0.45\textwidth]{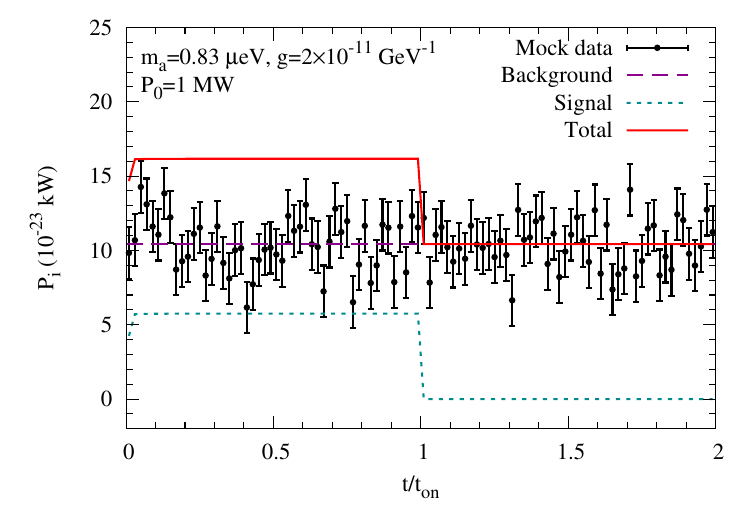}
\includegraphics[width=0.45\textwidth]{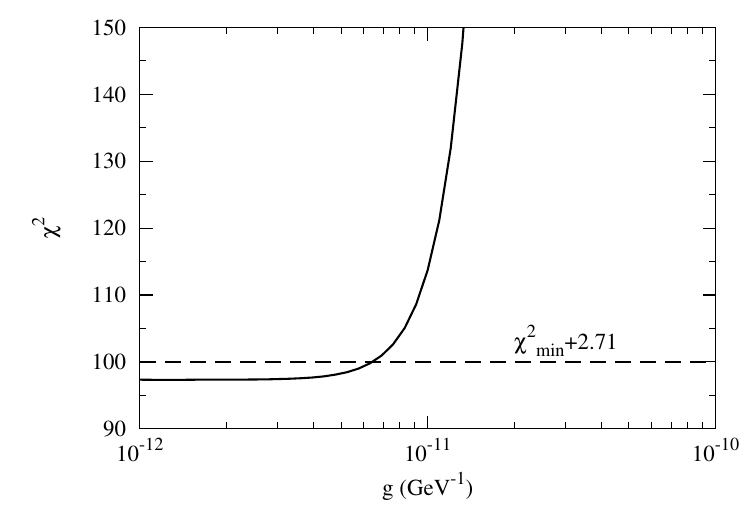}
\caption{Top: Simulated data for {\it pure background hypothesis}, folded for the whole 
3160 s of operation. The current 21CMA layout has been adopted. The horizontal 
dashed line is the average background, the blue dotted line corresponds to the expected 
echo signal for axions with $m_a=0.83~\mu$eV, $g=2\times10^{-11}$ GeV$^{-1}$, and an emitter 
power $P_0=1$ MW, and the red solid line is the sum of signal and background. 
Bottom: $\chi^2$ versus axion-photon coupling $g$. (Color online)}
\label{fig:simu}
\end{figure}

In the top panel of Fig.~\ref{fig:simu}, we show the simulated data for the pure background 
hypothesis and the standard 21CMA layout. 
For comparison, the expected echo signal from an axion model with 
$m_a=0.83~\mu$eV (corresponding to a frequency of axion-induced electromagnetic wave of 
100 MHz) and $g=2\times10^{-11}$ GeV$^{-1}$, and the sum of signal plus background are
shown by dotted and solid lines, respectively. Here we assume $P_0=1$ MW, $\ton=1$ s 
(to switch on the beam for $\ton$ and then to switch off for another $\ton$), with a time 
bin width of $\Delta t_i=0.02$ s. The integrated signal sending time is 1580 
s, and the total experiment time is 3160 s for one frequency band with 
$\delta\nu$ width. 
Obviously, the hypothetical axion signal exceeds the data significantly,
which means our sensitivity can reach an axion-photon coupling much lower than 
$g=2\times10^{-11}$ GeV$^{-1}$.

Varying the coupling constant $g$, we can get the $\chi^2$ distribution as shown in the 
bottom panel of Fig.~\ref{fig:simu}. Through setting the difference between $\chi^2$ for 
any $g$ and the minimum $\chi^2_{\rm min}$, $\Delta\chi^2=\chi^2-\chi^2_{\rm min}=2.71$, 
we can get the $95\%$ confidence level sensitivity on the axion-photon coupling for given 
mass $m_a$. For the above parameter setting, we get the $95\%$ sensitivity on $g$ to be
$6.5\times10^{-12}$ GeV$^{-1}$. For the optimized layout,
the signal will enhance by a factor of $\sim5.8$, and hence the sensitivity on $g$ will
improve by a factor of $\sqrt{5.8}=2.4$. We can further obtain the scaling relation between 
the sensitivity and the central frequency as $g=6.5\times10^{-12}(\nu/100~{\rm MHz})^{3/4}
(P_0/{\rm MW})^{-1/2}$ GeV$^{-1}$ for the standard 21CMA array, and 
$g=2.7\times10^{-12}(\nu/100~{\rm MHz})^{3/4}(P_0/{\rm MW})^{-1/2}$ GeV$^{-1}$ 
for the optimized array.

\end{document}